\documentclass[runningheads,a4paper]{llncs}

\usepackage{amssymb}
\usepackage{graphicx}
\usepackage{url}
\urldef{\mailparis}\path|{anastasios.bellas, marie.cottrell}@univ-paris1.fr|
\urldef{\maildescartes}\path|charles.bouveyron@parisdescartes.fr|
\urldef{\mailsnecma}\path|jerome.lacaille@snecma.fr|

\newcommand{\keywords}[1]{\par\addvspace\baselineskip
\noindent\keywordname\enspace\ignorespaces#1}
\usepackage{phd}
\begin{document}

\mainmatter  

\title{Anomaly detection  based on confidence intervals using SOM 
with an application to  Health Monitoring }

\titlerunning{Anomaly detection}

%
%
\author{Anastasios Bellas\inst{1}
\and Charles Bouveyron\inst{2}\and Marie Cottrell\inst{1}\and Jerome Lacaille\inst{3}}
\authorrunning{Anomaly detection}

\institute{SAMM, Universit\'e Paris 1 Panth\'e on-Sorbonne\\
90 rue de Tolbiac, 75013 Paris, France\\
\mailparis\\ \and 
Laboratoire MAP5, Universit\'e Paris Descartes \& Sorbonne Paris Cit\'e\\
45 rue des Saints-Pères, 75006 Paris, France\\
\maildescartes\\ 
\and 
 SNECMA, Rond-Point Ren\'e Ravaud-R\'eau,\\
77550 Moissy-Cramayel CEDEX,  France \\
\mailsnecma }

%
%

\toctitle{Lecture Notes in Computer Science}
\tocauthor{Authors' Instructions}
\maketitle

\begin{abstract}
We develop an application of SOM for the task of anomaly detection and
visualization. To remove the effect of exogenous independent  variables, we use
a correction model which is more accurate than the usual one, since we apply
different  linear models in each cluster of context. We do not assume any
particular probability distribution of the data and the detection method is
based on the distance of new data to the Kohonen map learned with corrected
healthy data. We apply the proposed method to the detection of  aircraft engine
anomalies.

\keywords{Health Monitoring, aircraft, SOM, clustering, anomaly detection, confidence intervals}
\end{abstract}

\section{Introduction, Health monitoring and related works}\label{sec:intro}

In this paper, we develop SOM-based methods for the
task of anomaly detection and visualization of aircraft engine anomalies.

The paper is organized as follows : Section~\ref{sec:intro} is an 
 introduction to the subject, giving a small review of related
articles. In Section~\ref{sec:overview}, the different components of the system
proposed are being described in detail. Section~\ref{sec:data} presents the data
that we used in this application, the experiments that we carried out and their results. 
Section 4 presents a short conclusion.

\subsection{Health monitoring}

Health monitoring consists in a set of algorithms which monitor in real time the
operational parameters of the system.  The goal is to detect early signs of
failure, to schedule maintenance and to identify the causes of anomalies.

Here we consider a domain where Health Monitoring is especially important: 
aircraft engine safety and reliability. Snecma, the french aircraft engine
constructor, has developed well-established methodologies and innovative tools:
to ensure the operational reliability of engines and the availability of
aircraft, all flights are monitored. In this way, the availability of engines is
improved: operational events, such as D\&C (Delay and Cancellation) or IFSD
(In-flight Shut Down) are avoided and maintenance operations planning and costs
are optimized.

\subsection{Related work}

This paper follows other related works. For example, 
\cite{lacaille2010adaptive} have proposed the \emph{Continuous Empirical Score}
(CES), an algorithm for Health Monitoring for a test cell environment based on three
components: a clustering algorithm based on EM, a scoring component and a
decision procedure. \par

In~\cite{lacaille2009online,come2010aircraft,lacaille2011sudden}, a similar
methodology is applied to detect change-points in Aircraft Communication,
Addressing and Reporting System (ACARS) data, which are basically messages
transmitted from the aircraft to the ground containing on-flight measurements of
various quantities relative to the engine and the aircraft. \par

In ~\cite{come2010self}, a novel \emph{star} architecture for Kohonen maps
is proposed. The idea here is that the center of the star will capture the
normal state of an engine with some rays regrouping normal behaviors which have
drifted away from the center state and other rays capturing possible engine
defects.\par

In this paper, we propose a new  anomaly detection method, using statistical
methods such as projections on Kohonen maps and computation of confidence
intervals. It is adapted to  large sets of data samples, which are not
necessarily issued from a single engine. \par

Note that typically, methods for Health Monitoring use an extensive amount
of expert knowledge, whereas the proposed method is fully automatic and has not
been designed for a specific dataset.\par

Finally, let us note that the reader can find a broad survey of methods for
anomaly detection and their applications in~\cite{chandola2009outlier}
and \cite{Markou2003a,Markou2003b}.

\section{Overview of the methodology}\label{sec:overview}

Flight data consist of a series of measures acquired by sensors
positioned on the engine or the body of the aircraft. Data may be issued from a
single or multiple engines. We distinguish between \textit{exogenous} or
\textit{environmental} measures related to the environment and
\textit{endogenous} or \textit{operational} variables related to the engine
itself. The reader can find the list of variables in Table~\ref{tab:variables}. 
For the anomaly detection task, we are interested in operational
measures. However, environmental influence on the operational measures needs to
be removed to get reliable detection. 

\begin{table}[t]
\centering
\begin{tabular}{|l|l|}
  \hline
 \textbf{Name} & \textbf{Description}\\
  \hline
  \multicolumn{2}{|l|}{\textbf{Operational variables}} \\
 \hline
  EXH & Exhaustion gas temperature \\
  \hline
  N2 & Core speed \\
  \hline
  Temp1 & Temperature at the entrance of the fan \\
  \hline
  Pres & Static pressure before combustion \\
  \hline
  Temp2 & Temperature before combustion \\
  \hline
  FF & Fuel flow \\
  \hline
  \multicolumn{2}{|l|}{\textbf{Environmental variables}} \\
 \hline
  ALT & Altitude \\
  \hline
  Temp3 & Ambient temperature \\
  \hline
  SP & Aircraft speed \\
  \hline
  N1 & Fan speed \\
  \hline
  \multicolumn{2}{|l|}{\textbf{Other variables}} \\
  \hline
  ENG & Engine index \\
  \hline 
  AGE & Engine age\\
  \hline
 \end{tabular}
\caption{Description of the variables of the cruise phase data.}
\label{tab:variables}
\end{table}

The entire procedure consists of two main phases. 

\begin{enumerate}
\item The first phase is the \textit{training} or \textit{learning} phase where
we learn based on healthy data.

\begin{itemize}
\item We cluster data into clusters of environmental
conditions using only environmental variables. 

\item We correct operational measures variables from the influence of the
environment using a linear model, and we get the residuals (corrected values). 

\item Next, a SOM is being learned based on the residuals. 

\item We calibrate the anomaly detection component by computing the
confidence intervals of the distances of the corrected data to the SOM.
\end{itemize}

\item The learning phase is followed by the \textit{test} phase, where
novel data are taken into account.

 \begin{itemize}
\item Each novel data sample is being clustered in one of the
environment clusters established in the training phase.

\item It is then being corrected of the environment influence  using the
linear model estimated earlier.  

\item The test sample is projected to the Kohonen map constructed in the
training phase and finally, the calibrated anomaly detection component
determines if the sample is normal or not. \par

\end{itemize}

\end{enumerate}


\subsubsection{Clustering of the environmental contexts}

An important point is the choice of the clustering method. Note that clustering
is carried out on the \textit{environmental}
variables. The most popular clustering method is the Hierarchical Ascending
Classification \cite{CAH} algorithm, which allows us to choose the number of
clusters based on the explained variance at different heights of the constructed
tree.\par

However in this work our goal is to develop a more general methodology that
could process even high-dimensional data and it is well-known
that HAC is not adapted to this kind of data. Consequently, we are
particularly interested in methods based on subspaces such as
HDDC~\cite{bouveyron2007high}, since they can provide us with a
parsimonious representation of high-dimensional data. Thus, we will use HDDC for
the environment clustering, despite its less good performance for
low-dimensional data.

\subsubsection{Corrupting data}\label{sec:anomalies}

\begin{figure}[t]
\centering
 \subfloat[]{\label{fig:anomaly_example-a}\includegraphics
[width=0.5\columnwidth]
{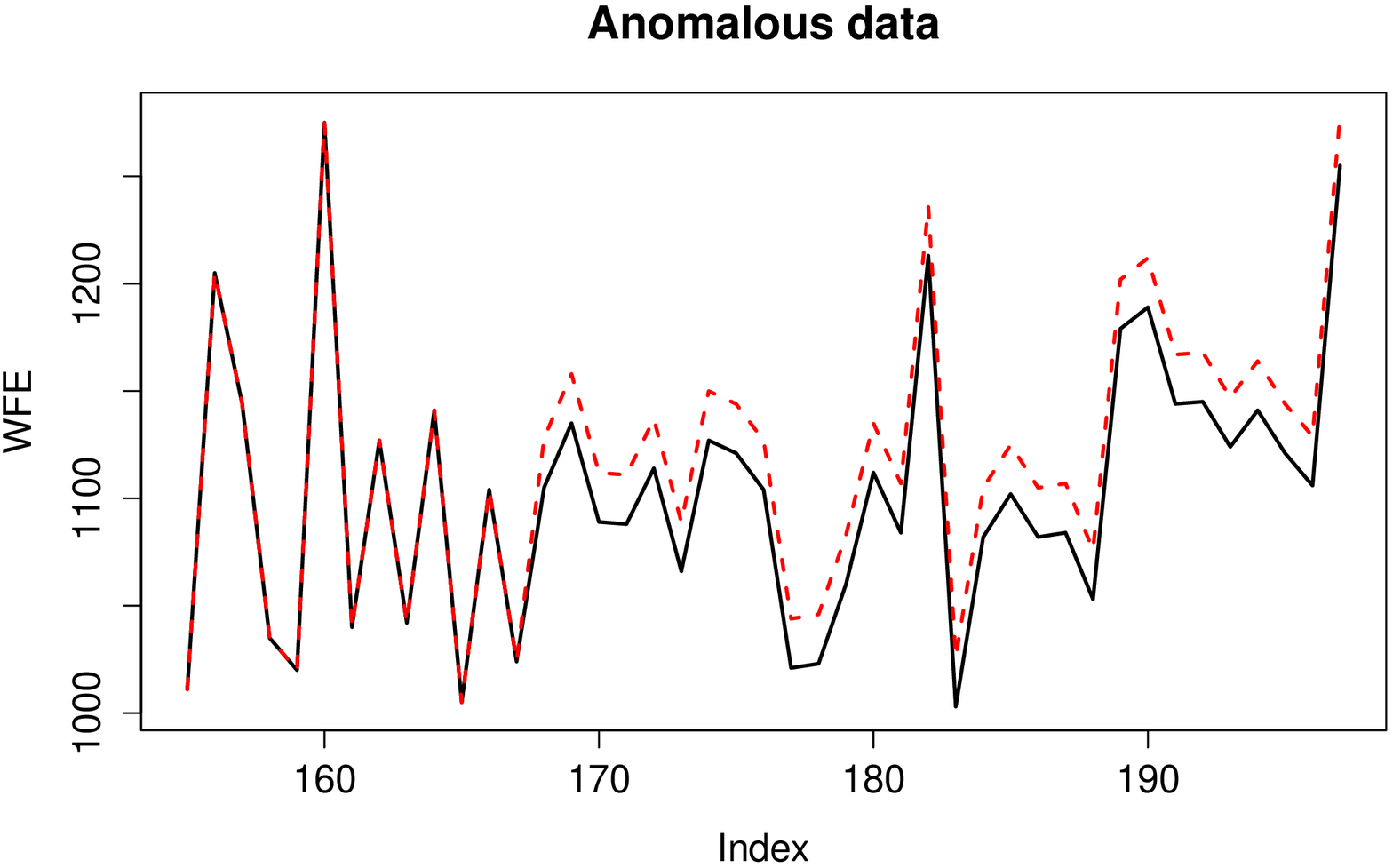}}
 \subfloat[]{\label{fig:anomaly_example-b}\includegraphics[width=0.5\columnwidth]
{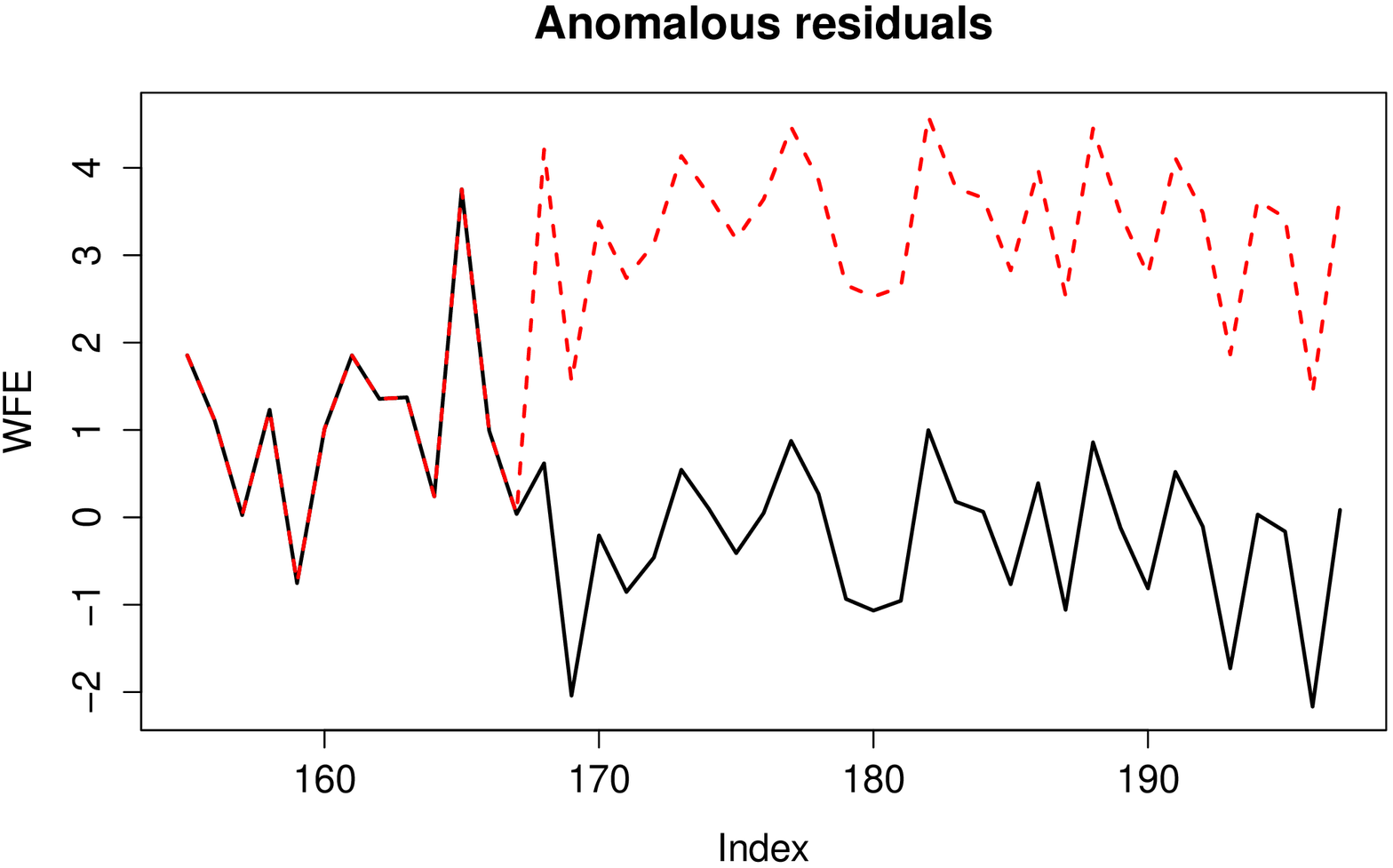}}
\caption{An example of an anomaly of the FF variable of the cruise flight
data (a) Superposition of the healthy data (solid black lines) and the data
with anomalies (dashed red line) (b) Superposition of the corrected data obtained from the 
healthy data and corrected data obtained from corrupted data. 
The anomaly is visible only on corrected data.}
\label{fig:anomaly_example}
\end{figure}

In order to test the capacity of the proposed system to detect anomalies, we
need data with anomalies. However, it is very difficult to get them due to the
extraordinary reliability of the aircraft engines and we cannot
fabricate them because deliberately damaging the engine or the test cell is
clearly not an option. Therefore, we create artificial anomalies by
corrupting some of the data based on expert specifications that have been
established following well-known possible malfunctions of aircraft engines. \par

Corrupting the data with anomalies is carried out according to a
\emph{signature} describing the defect (malfunction). A signature is a
vector $\mathbf{s} \in \mathbb{R}^p$. Following $s$, a corruption term is added
to the nominal value of the signal for a randomly chosen set of successive data
samples. \par

Figure~\ref{fig:anomaly_example-a} gives an example of the corruption of the FF
variable for one of the engines. Figure~\ref{fig:anomaly_example-b} shows the
corrupted variable of the corrected data, that is, after having removed the
influence of the environmental variables.

\subsection{Clustering the corrected data using a SOM}

In order to build an anomaly detection component, we need a clustering method to
define homogeneous subsets of corrected data. We choose to use the 
SOM algorithm~\cite{kohonen2001self} for its well-known properties of clustering
organized with respect to each variable of the data as well as its
visualization ability.


The output of the algorithm is a set of prototype vectors that define an
"organized" map, that is, a map that respects the topology of the data in the
input space. We can then color the map according to the distribution of the data
for each variable. In this way, we can visually detect regions in the map where
low or high values of a given variable are located. A smooth coloring shows
that it is well organized. In the next section, we show how
to use these properties for the anomaly detection task.

\subsection{Anomaly detection}

In this subsection, we present two anomaly detection methods that are based on
confidence intervals. These intervals provide us with a "normality" interval of
healthy data, which we can then use in the test phase to determine if a novel
data sample is healthy or not. \par

We have already seen that the SOM algorithm associates each data sample with
the nearest prototype vector, given a selected distance measure. Usually, the
Euclidean distance is selected. Let $L$ be the number of the units of the map,
$\{m_l,\text{ } l=1, \ldots, L\}$ the prototypes. For each data
sample, we calculate $\mathbf{x}_i$, its distance to the map, namely the
distance to its nearest prototype vector:
\begin{eqnarray}
 d(\mathbf{x}_i) & = & \min\limits_l \norm{\mathbf{x}_i - \mathbf{m}_l}{}^2
\label{eq:dist}
\end{eqnarray}
where $i=1,\dotsc,n$. Note that this way of calculating distance will give us a
far more useful measure than if we had just utilized the distance to the global
mean, \emph{i.e.} $d(\mathbf{x}_i) = \norm{\mathbf{x}_i -
\mathbf{\bar{x}}}{}^2$.\par

The confidence intervals that we use here are calculated using
distances of training data to the map. The main idea is that the distance
of a data sample to its prototype vector has to  be "small". So, a "large"
distance could possibly indicate an anomaly. We propose a global and a local variant of this method.

\subsubsection{Global detection}

During the training phase, we calculate the distances 
$d(\mathbf{x}_i)$, $\forall i$, according to Equation (1). We can thus construct a
confidence interval by taking the $99$-th percentile of the distances,
$P_{99}(\{d(\mathbf{x}_i),~\forall i\})$, as the upper limit. The lower limit is
equal to $0$ since a distance is strictly positive. We define thus the
confidence interval~$\mathcal{I}$
\begin{eqnarray}
 \mathcal{I} &=& \left[ 0, P_{99}(\{d(\mathbf{x}_i),~\forall i\}) \right]
\label{eq:global_confint}
\end{eqnarray}
For a novel data sample $\xx$, we establish the following decision rule:
\small
\begin{eqnarray}\label{eq:decision_global}
 \left\{
  \begin{array}{l l}
    \text{The novel data sample is healthy, if }
d(\xx) \in \mathcal{I} \\
    \text{The novel data sample is an anomaly, if }
d(\xx) \notin \mathcal{I}. \end{array} \right.
\end{eqnarray}
\normalsize
The choice of the $99$-th percentile is a compromise taking into account our 
double-sided objective of a high anomaly detection rate with the smallest
possible false alarm rate. Moreover, since the true anomaly rate is typically
very small in civil aircraft engines, the choice of such a high percentile,
which also serves as an upper bound of the normal functioning interval, is
reasonable.

\subsubsection{Local Detection}

In a similar manner, in the training phase, we can build a confidence interval
for every cluster $l$. In this way, we obtain $L$ confidence intervals
$\mathcal{I}_l$, $l=1, \dotsc, L$ by taking the $99$-th percentile of the
\emph{per} cluster distances as the upper limit
\begin{eqnarray} 
 \mathcal{I}_l &=& \left[ 0, P_{99}\left( \{d(\mathbf{x}_i): \mathbf{x}_i
\text{ in SOM cluster } l
\}\right)
\right]
\label{eq:local_confint}
\end{eqnarray}
For a novel data sample $\xx$ (in the test phase), we establish the following
decision rule:
\begin{eqnarray}\label{eq:decision_local}
 \left\{
  \begin{array}{l l}
    \text{The novel data sample, affected to SOM cluster $l$, is healthy, if }
d(\xx)
\in \mathcal{I}_l \\
    \text{The novel data sample, affected to SOM cluster $l$, is an anomaly if }
d(\mathbf{x}) \notin \mathcal{I}_l. \end{array} \right.
\end{eqnarray}

\section{Application to aircraft flight cruise data}\label{sec:data}
In this section, we present the data that we used for our experiments as well as
the  processing that we carried out on them.

Data samples in this dataset are snapshots taken from the cruise phase of a
flight. Each data sample is a vector of endogenous and environmental variables,
as well as categorical variables. Data are issued from $16$ distinct
engines of the same type. For each time instant, there are two
snapshots, one for the engine on the left and another one for the engine on the
right. Thus, engines appear always in pairs. Snapshots are issued from
different flights. Typically, there is one pair of snapshots per flight. 
The reader can find the list of variables in Table~\ref{tab:variables}. The
dataset we used here contains $2472$ data samples and $12$ variables.\par 

We have divided the dataset into a training set
and a test set. For the training set, we randomly picked $n=2000$ data
samples among the  $2472$ that we dispose of in total. The test set is composed
of the $472$ remaining data samples.  We have verified that all engines are
 represented in both sets.  We have sorted data based on the engine ID
(primary key of the sort) and for a given engine, based on the timestamp of the
snapshot. We normalize the data (center and scale) because the
scales of the variables were very different.

\subsubsection{Selection of the number of clusters in environment clustering}
Clustering is carried out on environmental variables to define
clusters of contexts. Due to the large variability of the different contexts
(extreme temperatures very high or very cold and so on), we have to do a
compromise between a good variance explanation and a reasonable number of
clusters (to keep a sufficient number of data in each cluster).  If we compare
HDDC
to the Hierarchical Ascending Classification (HAC) algorithm in terms of
explained variance, we observe that the explained variance is about 50 \% for
five clusters for both algorithms. And as mentioned before, we prefer to use
HDDC~\cite{bouveyron2007high} to present a methodology which can be easily
adapted to high-dimensional data. Let $K=5$ be the number of clusters.

\subsubsection{Correcting the endogenous data from environmental influence}

We correct the operational variables of environmental influence using the
procedure we described in section 2. After the partition into 5 clusters
based on environmental variables, we compute the residuals of the operational
variables as follows: if we  set  
$X^{(1)} =$ N1, $X^{(2)} =$ Temp3, $X^{(3)} =$ SP, $X^{(4)} = $ ALT et $X^{(5)}
= $ AGE, we write
\begin{align}
 \notag Y_{rkj}  & = \mu + \alpha_r  + \beta_k  +
\gamma_{1k}
X^{(1)}_{rkj}  + \gamma_{2k} X^{(2)}_{rkj} + \gamma_{3k} X^{(3)}_{rkj} + \\
&  \gamma_{4k} X^{(4)}_{rkj} + \gamma_{5} X^{(5)}_{rkj} + \varepsilon_{rkj}
\label{eq:lm}
\end{align}
where $Y$ is one of the $d=6$ operational variables, $r \in \{1,\dotsc,16\}$ is
the engine index, $k \in \{1, \dotsc, 5\}$ is
the cluster number, $j \in  \{1, \dotsc, n_{rk}\}$ is the observation index.
Moreover, $\mu$ is the intercept, $\alpha_r$ is the effect of the engine and
$\beta_k$ the effect of the cluster.

\subsubsection{Learning a SOM with residuals}
 
By analyzing the residuals, one can observe  that the model succeeds in
capturing the influence of the environment on the endogenous measures, since the
magnitude of the residuals is rather small (between -0.5 and + 0.5). The
residuals therefore capture behaviors of the engine which are not due to
environmental conditions. The residuals are expected to be centered, \emph{i.e.}
to have a mean equal to $0$. However, they are not necessarily scaled, so we
re-scale them. \par 

\begin{figure}[t]
 \centering
\includegraphics[width=1.2\textwidth]{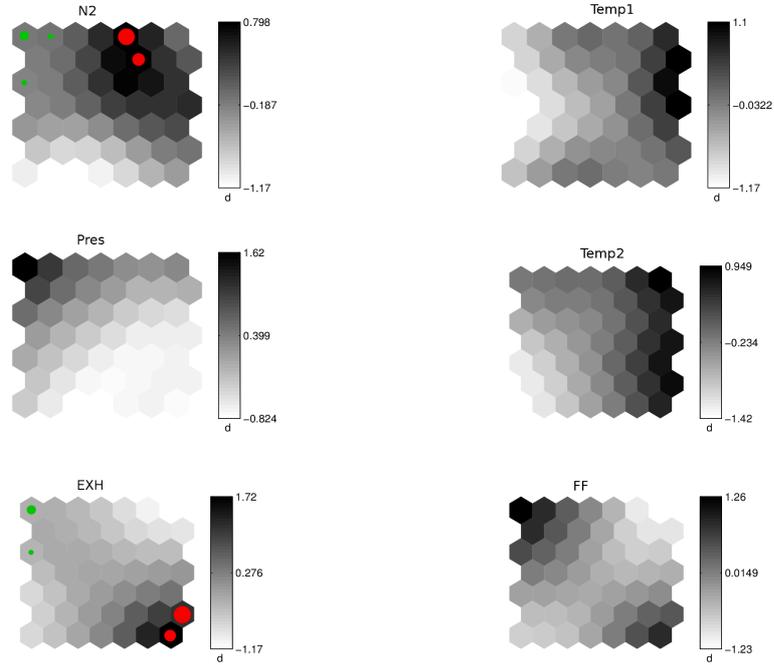}
\caption{SOM built from the corrected training residuals for each of the $p=6$
endogenous variables. Black cells contain high values of the variable while
white ones contain low values. Red dots refer to anomalies and green dots
to healthy data for two different types of defects bearing on the
variables N2 and EXH. The proposed method clusters them in different regions of
the map. The size of each dot is proportional to the number of points of the
cluster.}
\label{fig:som_app}
\end{figure}

Generally speaking, since residuals are not smooth, we carry out smoothing
using a moving average of width $w=7$ (central element plus $3$ elements on
the left plus $3$ elements on the right). We note that by smoothing, we lose
$\lfloor \frac{w}{2} \rfloor$ data samples from the beginning and the end.
Therefore, we end up with a set of $1994$ residual samples instead
of the $2000$ that we had initially. Next, we construct a Self-Organizing Map
(SOM) based on the residuals (Figure~\ref{fig:som_app}). We have opted here
for a map of $49$ neurons ($7 \times 7$) because we need a minimum of
observations per SOM cluster in order to calculate the normal
functioning intervals with precision. \par

The last step is the calibration of the detection component by determining the global
and local confidence intervals based on the distances of the data to the map. 
For the global case, according to Equation~\ref{eq:global_confint}, we have:
\begin{eqnarray*}
 \mathcal{I} & = & \left[0, 4.1707 \right]
\end{eqnarray*}
In a similar manner, we derive the upper limits of the local confidence
intervals, ranging from $1.48$ to $6.03$.
%

\subsubsection{Test phase}

In the test phase, we assume that novel data samples are being made available.
We first corrupt these data following the technique proposed in
Section~\ref{sec:anomalies}. Snecma experts provided us with signatures of $12$
known defects (anomalies), that we added to the data. For data confidentiality
reasons, we are obliged to anonymize the defects and we refer to them as "Defect
$1$", "Defect $2$" etc.\par 

We start by normalizing test data with the coefficients used
to normalize training data earlier. We then cluster data into environment
clusters using the model parameters we estimated on the training data earlier. 
Next, we correct data from environmental influence using the model we built on 
the training data. In this way, we obtain the test residuals, 
that we re-scale with the same scaling coefficients used to
re-scale training residuals. \par 

We apply a smoothing transformation using a moving average, exactly like we did
for training residuals. We use the same window size, \emph{i.e.} $w=7$.
Smoothing causes some of the data to be lost, so we end up with $466$ test
residuals instead of the $472$ we had initially. \par

Finally, we project data onto the Kohonen map that we built in the training
phase and we compute the distances $d(\mathbf{x})$ as in equation (1). 
We apply the decision rule, either the global decision rule
of~(\ref{eq:decision_global}) or the local one of~(\ref{eq:decision_local}).
\par

 In order to evaluate our system, we calculate the detection rate ($tpr$) and
the false alarms rate ($pfa$):
\begin{eqnarray*}
tpr &=& \frac{\text{number of detections}}{\text{number
of anomalies}}\\ \\
pfa &=& \frac{\text{number of non-expected detections}}
{\text{number of detections}} \\
\end{eqnarray*}
\begin{table}[t]
\small
\centering
 \begin{tabular}{|l|ll|ll|}
  \hline
  & \multicolumn{2}{|l|}{\textbf{Global detection}} &
\multicolumn{2}{|l|}{\textbf{Local detection}} \\
\hline
  \textbf{Defect} & $\mathbf{tpr}$ & $\mathbf{pfa}$  &
$\mathbf{tpr}$ & $\mathbf{pfa}$   \\
  \hline
  Defect $1$ & 100\% & 18,9\%  & 100\% & 45,4\% \\
  \hline
  Defect $2$ & 100\% & 11,4\%  & 100\% & 42,6\%  \\
  \hline
  Defect $3$ & 100\% & 16,7\%  & 100\% & 47,9\%  \\
  \hline
  Defect $4$ & 100\% & 15,1\%  & 100\% & 45,1\%  \\
  \hline
  Defect $5$ & 96,7\% & 14,7\%  & 100\% & 43,4\%  \\
  \hline
  Defect $6$ & 100\% & 13,9\%  &  100\% & 43,6\% \\
  \hline
  Defect $7$ & 96,7\% & 12,1\% & 96,7\% & 44,2\% \\
  \hline
  Defect $8$ & 100\% & 26,3\%  & 100\% & 50\%  \\
  \hline
  Defect $9$ & 100\% & 15,8\%  & 100\% & 43,9\% \\
  \hline
  Defect $10$ & 100\% & 26,7\% & 100\% & 55,1\%  \\
  \hline
  Defect $11$ & 100\% & 17,1\%& 100\% & 46,3\%  \\
  \hline
  Defect $12$ & 100\% & 21\% & 100\% & 46,4\% \\
  \hline
 \end{tabular}
\caption{Detection rate ($tpr$) and false alarm rate ($pfa$) for different types
of defects and for both anomaly detection methods (global and local) for test data.}
\label{tab:taux}
\end{table}
In Table~\ref{tab:taux}, we can see detection results for all $12$ defects and
for both detection methods (global and local). It is clear that both methods
succeed in detecting the defects, almost without a single miss. 
The global method has a lower false alarm rate than the local one. 
This is because in our example, confidence intervals cannot be calculated 
reliably in the local case since we have few data per SOM cluster.\par

\begin{figure}[h!t]
\includegraphics[width=\columnwidth,height=11cm]
{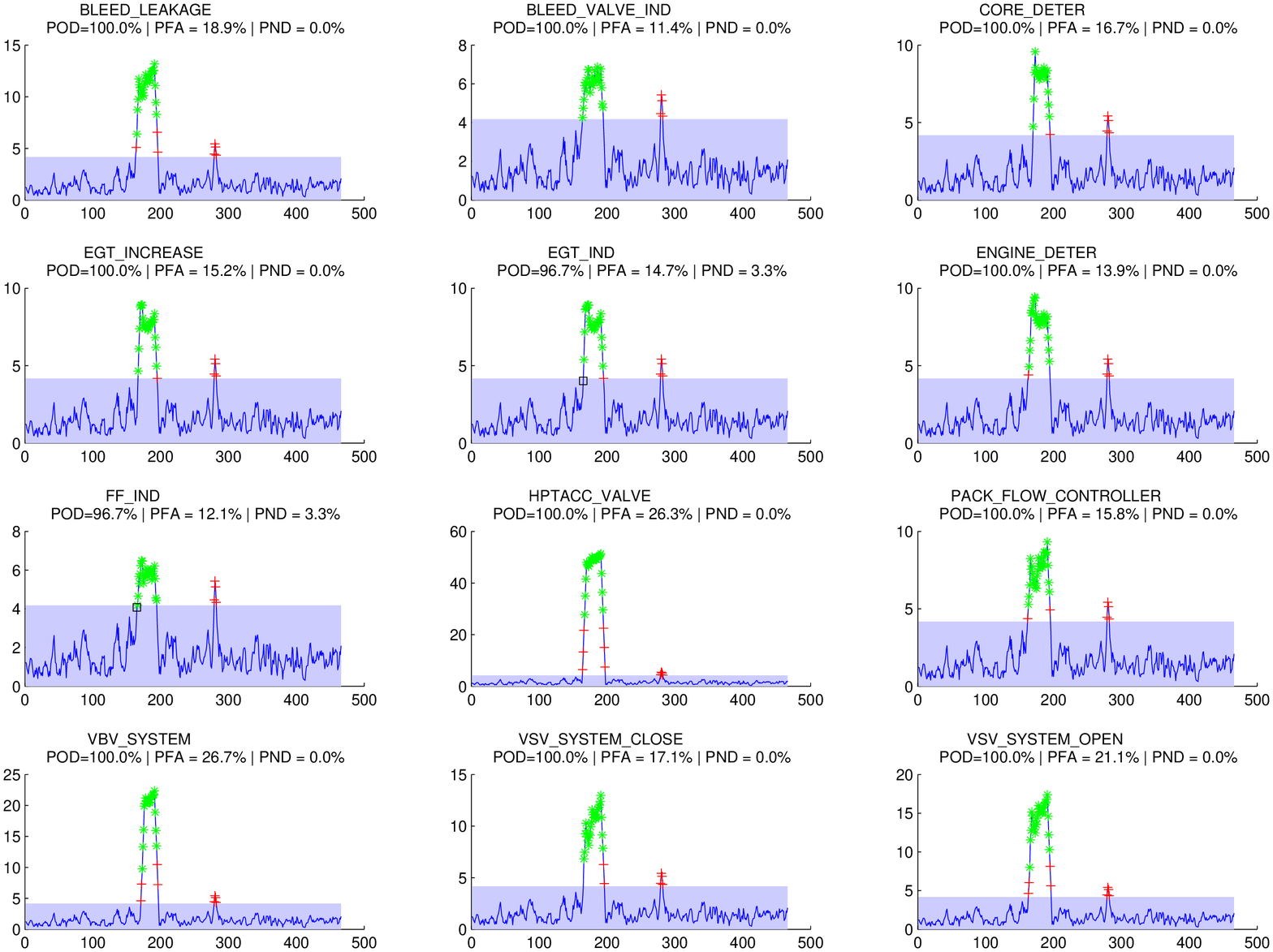}
\caption{Distances of the test data to their nearest prototype vector and
the global confidence interval (in light blue). Red crosses show the false
alarms and green stars show successful detection.}
\label{fig:gConfint}
\end{figure}

Figure~\ref{fig:gConfint} shows the distance $d$ of each data sample
(samples on the horizontal axis) to their nearest prototype vector (Equation
~\ref{eq:dist}). The light blue band shows the global confidence
interval $\mathcal{I}$ that we calculated in the training phase. Red
crosses show the false alarms and green stars the correct
detections.
%
%
%
%
Due to limited space in this contribution, the figures related to the local detection
can be found in the following URL:
https://drive.google.com/folderview?id=0B0EJciu-PLatZzdqR25oVjNNaTg\&usp=sharing
\section{Conclusion and Future work}
We have developed an integrated methodology for the analysis,
detection and visualization of anomalies of aircraft engines. We have developed a
statistical technique that builds intervals of "normal" functioning of an engine based on distances of
healthy data from the map with the aim of detecting anomalies. The system is
first calibrated using healthy data. It is then fully operational and can
process data that was not seen during training.\par

The proposed method has shown satisfying performance in anomaly detection, given
that it is a general method which does not incorporate any expert knowledge and
that it is, thus, a general tool that can be used to detect anomalies in
any kind of data.\par

Another advantage of the proposed method is that the use
of the dimension allows to carry out multi-dimensional anomaly detection in a
problem of dimension $1$. Moreover, the representation of the operational
variables given by the use of the distance to the SOM is of a higher
granularity than that of the distance from the global mean. Last but not least,
the use of SOM allows us to give interesting visualizations of healthy and
abnormal data, as seen in Figure~\ref{fig:som_app}. \par

An extension of our work would be to carry out anomaly detection for
datastreams using this method. A naive solution would be to re-calibrate the
components of the system with each novel data sample, but it would be very
time-consuming. Instead, one can try to make each component of the system to
operate on datastreams.

\end{document}